%% file: main.tex
\documentclass[journal]{IEEEtran}

\usepackage{amsmath}

\usepackage{amssymb}
\usepackage{amsthm}
\usepackage{amsxtra}
\usepackage{amsfonts}
\usepackage{graphicx}
\usepackage{latexsym}
\usepackage{epsfig}
\usepackage{verbatim}
\usepackage{bm}
\usepackage{bbm}
\usepackage{color}
\usepackage{varwidth}
\usepackage{subfigure}
\usepackage{wrapfig}
\usepackage{trivfloat}
\usepackage{todonotes}
\usepackage{url}
\usepackage{fullpage}
\usepackage{dsfont}
\usepackage[margin=9pt,font=small,labelfont=bf,labelsep=colon]{caption}
\usepackage{algorithm}
\usepackage{algpseudocode}
\usepackage{pifont}

\usepackage{lipsum}
\usepackage[procnames]{listings}
\usepackage{array}
\usepackage{multicol}
\usepackage{float}
\usepackage{url}
\usepackage{subfigure}
\usepackage{multirow}
\usepackage{enumitem}
\usepackage{amsmath,lipsum}

\usepackage[subtle]{savetrees}

\newtheorem{theorem}{Theorem}

\floatname{algorithm}{Procedure}

\onecolumn
\title{Network Topology Mapping from Partial Virtual Coordinates and Graph
Geodesics}
\date{\today}
\setlist[itemize]{leftmargin=0.12in}

\begin{document}

\author{
  Anura~P.~Jayasumana,~Randy~Paffenroth,~Gunjan~Mahindre,~Sridhar~Ramasamy,~and~
  Kelum~Gajamannage

\thanks{A. P. Jayasumana, S. Ramasamy and G. Mahindre are with the Department of
  Electrical \& Computer Engineering, Colorado State University, Fort
  Collins, CO 80525}%
\thanks{R. Paffenroth is with the Department of Mathematical Sciences,
  Department of Computer Science, and the Data Science Program,
  Worcester Polytechnic Institute, Worcester, MA 01609}
\thanks{K. Gajamannage is with the Department of Mathematical Sciences,
 Worcester Polytechnic Institute, Worcester, MA 01609}
}

\maketitle
\begin{abstract}
For many important network types (e.g., sensor networks in complex harsh
environments and social networks) physical coordinate systems (e.g.,
Cartesian), and physical distances (e.g., Euclidean), are either difficult to
discern or inapplicable.  Accordingly, coordinate systems and characterizations based on
hop-distance measurements, such as Topology Preserving Maps (TPMs)
and Virtual-Coordinate (VC) systems 
are attractive alternatives to Cartesian coordinates for
many network algorithms. Herein,   we present an approach
to recover geometric and topological properties  of a network
with a small set of distance measurements.  In particular, our
approach is a combination of shortest path (often called geodesic) recovery
concepts and low-rank matrix completion, generalized to the case of
hop-distances in graphs.
Results for sensor networks embedded in 2-D and 3-D spaces, as well as a social networks, indicates that the method can accurately capture the network connectivity with a small set of measurements. 
TPM  generation can now also be based
on various context appropriate measurements or VC systems, as long as they
characterize different nodes by distances to small sets of random nodes (instead
of a set of global anchors).  The proposed method is a significant
generalization that  allows the topology to be extracted from a random set of
graph shortest paths, making it applicable in contexts such as social networks
where VC generation may not be possible.

\end{abstract}


\input{introduction}
\input{theory}
\input{approach}
\input{results}

\input{conclusions}
\small
\bibliographystyle{plain}

\input{appendix}

\end{document}

%% file: introduction.tex
\section{Introduction}

Large and complex networks naturally arise in communication and social
networks, the Internet-of-Things (IoT), and many other systems of importance.  
Data embedded in such networks exhibit distributed,
intricate, and  dynamic patterns. Our ability to extract information from and about these networks, detect anomalies,  and influence their performance can be substantially improved by a deeper understanding of their local and global structures \cite{Zhang2017,Zhang2015a,53}. Yet such operations are  
often impeded due to 
the size and complexity of these networks, and 
constraints such as energy,  measurement cost, and accessibility that 
prevent \emph{geometric and topological} structure of the
entire network from being fully observed, measured, characterized, or processed.
Inferences have to be made and patterns should be detected in the absence of
complete information. 
Accordingly, in this paper, we derive and demonstrate
novel techniques for detecting and representing network structures, e.g., topology, connectivity, and layout,  which are
scalable in their computation and communication, as well as graceful in their
degradation in the presence of limited measurements. In particular, we sample a network with a small set of  pair-wise hop-distances and use matrix completion techniques to capture the  topology of the network. We demonstrate the technique by deriving  topology preserving maps indicative of the layout of 2-D and 3-D sensor networks  with far fewer measurements compared to existing schemes. Furthermore, this new approach  extends applicability of hop-distance based techniques to cover complex networks such as social networks while providing a foundation for using a broader set of sampling strategies.    
\subsection{Motivating Examples}
\label{motivation}
Of specific interest to us are 
networks of inexpensive wireless devices such as smart Radio-Frequency Identification (RFID) tags or self-powered 
sensor nodes embedded in complex 2-D or 3-D surfaces and volumes. Here, the network features are
often characterized using the physical (Cartesian) coordinates of each node. One
then relies on the Euclidean properties of the network  layout (e.g., the
distance between nodes) for functions such as  area or volume coverage, topology
control, sensing, and routing. 
Physical location estimation is based on measurements such as the Received Signal Strength Indicator (RSSI) and time delay \cite{Le2017}. 
In networks deployed in harsh or complex environments, such methods are hampered or made completely ineffective  by issues such as
multi-path interference, reflections, shadowing and clock synchronization
\cite{oliveira2005error, Dhanapala2014}.  Although  physical coordinates provide a meaningful representation for node location,  for sensor networks deployed on complex surfaces and volumes, 
 \emph{graph distance} based techniques that use only connectivity measurements offer a more
robust, and attractive alternative \cite{Dhanapala2014}. For instance, in routing they overcome local minima caused by concave physical voids. Connectivity based techniques are also
generalizable to networks such as social networks, for which 
there is no notion of a location in a physical or Euclidean space. Even if a node is associated with a location, physical proximity does not necessarily imply connectivity. In fact for many such applications, what matters is the logical topology, i.e., connectivity and the hop distances among node pairs. 

\subsection{Hop-distance based Network Sampling}
\label{sampling}

We focus our attention on \emph{undirected graphs} $G$ defined by
$G=\left\{V, E\right\}$, where $V$ is the set of nodes or vertices and $E$ is a
set of edges corresponding to communication links, friendship status on a social
network, etc. 
Such a graph may be represented by
an \emph{adjacency matrix} $A$ \cite{Diestel2005}, where
\begin{equation}
A = \left[a_{ij}| a_{ij}=1 \;\text{if}\; (i,j) \in E,\  0
\;\text{otherwise}\; \right].
\label{A definition}
\end{equation}
Note that $a_{ij} = a_{ji}$ when $G$ is undirected. 
To make the current text concise, herein we focus on \emph{unweighted} graphs where
each $a_{ij} \in \{0,1\}$.  
%
Our main object of study will be \emph{Hop-Distance} Matrices (HDMs),  $H
\in \mathbb{N}^{N \times N}_0$, $\left( \text{where} \  \mathbb{N}_0 \right.$ \  denotes the non-negative integers $\left.\mathbb{N}\cup\{0\}\right)$, where
\begin{equation}
\begin{aligned}
h_{ij} = & \ \text{the length of the shortest path} \\
& \ \text{from node} \ i \ \text{to node} \ j .
\end{aligned}
\label{H definition}
\end{equation}
\noindent 
$H$, by construction, is symmetric and is invariant to the
situation where two nodes have multiple paths between them of the same shortest
length.  
From a mathematical perspective, hop-distances $h_{ij}$ may be thought of as the
computation of \emph{geodesics} (or shortest paths) in a graph
\cite{Diestel2005,Tenenbaum2000,Lee2007}, a perspective that we will make use of in the
sequel.
In particular, herein the term \emph{geodesic} will refer to a shortest
path between nodes, either when embedded in a particular space (e.g., the sum of 
straight line distances between node pairs forming a path in a Euclidean space), or when considered
as a node in a graph (e.g., as computed using Dijkstra's algorithm
\cite{Mehlhorn2008}).

Our goal is to capture the topology accurately using efficient network sampling schemes.   
Prominent among hop-distance based sampling methods are those relying on anchor-based
Virtual Coordinate Systems (VCS), in which each node is represented by a Virtual
Coordinate (VC) vector corresponding to the \emph{minimum} hop-distances to a
small set of nodes, known as \emph{anchor nodes}
\cite{Caruso2005,Liu2006,Rao2003}.
If we have $M$ anchor
nodes, then a VCS is an $N \times M$ sub-matrix $P$ of $H$ where the $i$-th row
provides an $M$ dimensional coordinate vector for the $i$-th node in the graph.
A VCS is then a ``relative'' coordinate system, as opposed to classic
``absolute'' coordinate systems such as Cartesian or Spherical.  In particular,
a VCS does not possess, or require, an absolute origin or other fixed geometric
properties such as a notion of angle.
However, a VCS lacks directional information as each coordinate indicates only
the distance to an anchor. Thus all the information about the network layout
such as shapes, voids, and boundaries are lost in a VCS. This issue leads to
natural generalizations of VCSs, such as Topology Coordinates (TCs) and Topology
Preserving Maps (TPMs) \cite{Dhanapala2014}, where one performs an
eigen-decomposition of $P$ or $H$ similar to that preformed in Principal Component
Analysis (PCA) \cite{Dhanapala2011f,Dhanapala2011g}.  Such methods can recover
the general shapes and boundaries of the physical network layout, thus providing
an effective alternative for representing 2-D and 3-D sensor networks and
carrying out operations such as routing and boundary detection, but without the
need for physical distance measurements \cite{Dhanapala2011f,Dhanapala2011g}.

\subsection{Contribution and Significance}
\label{novelty}
At a high-level, our approach is characterized by a) the choice of $H$ to represent a network,  b)
network sampling schemes that allow the measurement of  a small set of entries of $H$, e.g., to accommodate the constraints of accessibility, reduce the measurement complexity, etc., and c) filling in the incomplete sampled $H$ to obtain the network layout and topology.

A complete $H$ and a complete $A$ are interchangeable, i.e.,  to construct $A$ from $H$ one merely needs to set all entries with $h_{ij} > 1$ to $0$, while $H$ can
be obtained from  $A$ using procedures such as Dijkstra's algorithm
\cite{Mehlhorn2008}. However,  a \emph{partially observed $H$ and a
	partially observed $A$ are quite different}.  As opposed to an incomplete $A$,
we will demonstrate that an incomplete $H$ has quite interesting low-rank
properties, both from theoretical and practical perspectives, especially for
graphs arising from real-world networks.  Given the  properties such as low-rankness and an effective measurement scheme, the matrix $H$  can be
completed using efficient techniques for convex, low-rank matrix completion.
The newly completed $H$ can then be used to compute $A$ or to perform any other
desired analysis of the graph.

The main contribution of this paper is a technique that combines VC based techniques with low-rank matrix completion, that allows the extraction of topology and geometric features of a network from a small set of shortest path distances. 
\begin{itemize}
\item 
In case of networks embedded in 2-D and 3-D physical spaces, we
demonstrate that the topology preserving maps can be recovered using only a
small fraction of VC values, as opposed to existing method \cite{Dhanapala2014}
that requires the full set of VCs.
\item We broaden the possible set of network sampling schemes far beyond our previous results in 
\cite{Dhanapala2012,Dhanapala2011e,Dhanapala2014}, 
and point to a theory on which new sampling schemes may be grounded, i.e., selection of samples should not hinder the ability to complete the resulting incomplete low-rank matrices.   We demonstrate the reconstruction of the network topology from a small set of shortest path length measurements. Consequently, TPMs can now be generated based on a
variety of geodesic measurement approaches, of which an anchor-based VCS is one
instance.

\end{itemize} 
The novelty and the significance of this work may be gauged by the following impacts it can have on network analytics: 
\begin{itemize}
\item 
The theory of low-rank
matrix completion can now be applied for exploiting the sparseness of complex
real-world networks, and to develop communication and computation efficient
techniques for large-scale networks. 
Although 
$H$ for an arbitrary graph may not be low rank, we demonstrate that  for a broad class of complex real-life networks such distance matrices are
surprisingly low-rank.  
Our analysis is based upon ideas in low-rank
matrix completion
\cite{Lin2013,Candes2009,candes09ex,Paffenroth2013b,Paffenroth2012}, which have
been shown to scale to large problems with many thousands, if not millions, of
entries \cite{Paffenroth2013b,Paffenroth2012}, and the underlying topology that
we uncover is closely related to ideas in Non-linear Dimension Reduction (NDR), such
as Isomap \cite{Tenenbaum2000,Lee2007}, but generalized to the case where
only hop-distances are measured.

\item The 
applicability of hop-distance based property
extraction is extended to cases where certain distances and connectivities are not observable.
For large and complex networks, and especially those with access
limitations, we can likely not even measure a complete set of hop-distances to a set of common anchors. 
Examples of such cases include sensor and communication networks with restricted
access to certain nodes.
As explained in \cite{Beerliova2006} using examples from communication networks, it is realistic
to obtain the distances between nodes in many cases,
while it is difficult or impossible to obtain information
about edges or absence of edges that are far away
from the query node. The same argument extends
to many practical problems dealing with data, e.g.,
pathway prediction problem in proteins \cite{chou1990,Wang2005}, where the distance between two nodes
(proteins) can be evaluated yet finding the shortest
path by some measure (e.g., folding sequence) from
one node to the other is complex.
Social networks such as Facebook, Instagram, and  Orkut represent large scale networks. Controlled and  focused crawling \cite{Catan2011,Chak1999} are widely used approaches for data collection in such  networks for acquisition of data regarding hop distances, links, etc. As some profiles (nodes) are publicly accessible and some are not \cite{Misl2007}, we cannot gather data for all the nodes while crawling. In such cases, the hop distances acquired, or the links observed are only partial entries in the complete distance or adjacency matrix. 
Also, as proved in \cite{Koss2006}, missing data can have a significant impact on structural properties of a social network. 

\item Matrix completion algorithm finds many applications such as, finding top $N$ recommends for users \cite{MC1} and link prediction 
\cite{MC2}. While matrix completion as applied to Euclidean distance matrices is not new \cite{AlHomidan2005,Mishra2011,Krislock2012,Alfakih2000,Gower1982} our approach differs in a key aspect from those currently found in the literature. Our focus is on using \emph{hop-distances} rather than the more classic distance
measures such as Euclidean distances \cite{MC3}. In other words, instead of considering
Euclidean Distance Matrices (EDM) where each entry of the matrix codes for
the Euclidean distance between two nodes, we consider HDM where each entry of the matrix codes for the \emph{graph shortest path
distance} between two nodes. 

\item For a vast variety of networks, e.g., social and computer networks, 
\emph{the concept of Euclidean distances is not even applicable} (e.g., what is the ``Euclidean distance'' between two
people in a social network?). 
%
Accordingly, we leverage HDMs, which are the most
faithful representations of the network to which we have access.  As far as we
are aware, problems of such generality have not been considered before. 
\end{itemize}

Section~\ref{background} reviews the VC based sampling schemes. The theoretical considerations and results supporting proposed geodesic sampling schemes and low-rankness are outlined in Section~\ref{theory}. 
Methodology for VC based sampling and reconstruction is presented in Section~\ref{approach} followed by results in  Section~\ref{results} and conclusion. 
\section{Relation to Prior Work}
\label{background}
This section reviews hop-distance based network sampling schemes, the virtual-coordinate based representation, and their relationship to topology coordinates which recover the Euclidean properties of 2-D and 3-D networks.  
An anchor-based VCS is an M-dimensional
abstraction of the network connectivity, where each node is represented by an $M$ tuple, called VC vector which contains the  shortest path length
(in hops) from the node to each of a set of $M$ anchors
\cite{Cao2006,Caruso2005}. 
In an Aligned VC system \cite{Liu2006}, each node re-evaluates its coordinates by averaging its own coordinates with its neighbors' coordinates. Axis-based Virtual Coordinate Assignment Protocol \cite{Rao2003} estimates a 5-tuple VC for each node corresponding to longitude, latitude,
ripple, up, and down.   

A key question related to anchor-based VCS is the number $M$ and the placement
of anchors.  If an adequate number of anchors are not appropriately deployed, it
may cause the network to suffer from identical coordinates and local minima
\cite{Dhanapala2009CSR}, resulting in logical/virtual voids. The overhead
associated with  time and energy consumed for coordinate generation grows with
the number of anchors. The difficulty of
determining the optimal anchor set is compounded by the fact that the number of
anchors and their optimal placement are dependent on each other.  In Virtual Coordinate Assignment Protocol, the coordinates are defined
based on three anchors \cite{Caruso2005}, while all the perimeter nodes are selected in
\cite{Liu2006}.
Extreme Node Search uses two initial random anchors which allows the
selection of extreme nodes on internal and external boundaries as anchors
\cite{Dhanapala2011h}. 
An attractive alternative is to bypass anchor selection altogether, and select a set of random anchors.   Note, while from a networking perspective a random selection of
anchors may seem somewhat odd (i.e., a judicious placement might be viewed as
more appropriate), from a matrix completion perspective such a placement is not
only justified, but in many instances may be optimal
\cite{Lin2013,Candes2009,candes09ex,Paffenroth2012}. 
Again, such network organizational ideas have a dual perspective in the
mathematical literature, and the selection of anchors is closely related to the
idea of \emph{incoherence} in low-rank matrix recovery literature
\cite{Lin2013,Candes2009,candes09ex,Paffenroth2012}.


Geographical features such as boundaries and voids are missing from an anchor-based VCS. In our previous work \cite{Dhanapala2014,Jayasumana2016a}, recovery of geographic features from VCS is achieved by TPMs. TPMs derived from VCs  are maps that are nearly
homeomorphic to physical maps \cite{Dhanapala2014}. A TPM is a
distorted version of the real physical layout (map) in such a way that the
distortion accounts for connectivity information. In case of a 2D or 3D sensor
network with $M$ anchors, VCS is a mapping from the 2D or 3D network
layout to an $M$-dimensional space.  TPMs recover a 2D or 3D projection from this
$M$-dimensional representation such that it preserves the main features such as
boundaries and shapes of the network. Thus, TPMs can serve as an effective alternative
for physical coordinates in many network related functions.
Reconstruction of the graph from its  VCs is attempted in \cite{Ding2017} using an algorithmic approach. 
TC based schemes
have demonstrated performance comparable, or better than the corresponding
geographic coordinate based counterparts
\cite{Dhanapala2011f,Dhanapala2011g,Jayasumana2016a}.

While the algorithms that are based on VCs or TCs \cite{Dhanapala2014, Chatterjee2015} provide a viable, competitive and robust alternative to traditional geographic coordinate based methods, these techniques have so far required the complete set of VCs in order to extract the geometric information.  However, as we demonstrate here,
one can recover topological properties of a network \emph{without the need for
complete knowledge of the virtual coordinates}. 
While complexity of generating a complete distance matrix through flooding is of order $O(N^2)$, that for VC generation from a set of $M (M<<N)$ anchors is only $O(NM)$.
Herein, we further reduce this computational cost and
justify those results. In particular, we demonstrate a connection between TPMs,
NDR (Nonlinear Dimension Reduction) \cite{Tenenbaum2000,Lee2007}, and low-rank matrix
completion problems
\cite{Lin2013,Candes2009,candes09ex,Paffenroth2013b,Paffenroth2012}.

Finally, we observe that the computation of a low-rank approximation of an EDM using PCA
is equivalent to the Multi-Dimensional Scaling (MDS) \cite{Kruskal1964,Borg2005}
algorithm.  Even closer to our proposed technique, many authors consider
\emph{geodesic} adjacency matrices $A$ \cite{Tenenbaum2000,Lee2007} generated by
drawing short range, or \emph{neighbor}, distances from the EDM ($D$). They then
compute \emph{long range} distances by way of \emph{shortest path distances in
the graph represented by $A$}.  A low rank approximation of such a graph shortest
path based distance matrix is equivalent to the Isomap
\cite{Tenenbaum2000,Lee2007} algorithm for NDR.

%% file: theory.tex
\section{Theoretical Considerations}
\label{theory}


Herein, we are interested in studying the low-rank structure of \emph{graphs}
that arise in real-world network applications and we begin 
by
presenting a number of important theoretical ideas in this domain.   In
particular, we focus on 
two key questions when deriving our graph sampling and
reconstruction schemes.

\begin{itemize}

  \item First, what is the appropriate type of measurement to make?  For
  example, it is quite classic to represent a graph as its \emph{adjacency}
  matrix. However, we propose matrix completion is more effective when starting
  from a graph's \emph{distance} matrix,  even though the two representations
  are commonly viewed as being equivalent.
  As
  opposed to the adjacency matrix $A$, the hop distance matrix $H$ provides
  \emph{global information} about the graph.  In particular, each entry of $H$
  provides constraints on \emph{many entries of $A$}, with the simplest example of
  such global information being the Triangle-Inequality \cite{Abramowitz1964}.

  \item Second, what are the appropriate structural assumptions to make for reconstruction? 
  Given a partially observed adjacency matrix, any completion is perfectly consistent
  since  an adjacency matrix only provides local information.  Even a
  distance matrix can only provide bounds (e.g., the triangle inequality).  Accordingly,
  an appropriate structural assumption must be made (e.g., low rankness of $H$).
\end{itemize}

\subsection{Graph reconstruction:  Adjacency Matrices}

Given partial knowledge of a graph, there are several techniques 
for predicting the unknown information about the graph, including
an array of combinatorial techniques.  Examples include NP-hard combinatorial
algorithms such as minimal Hamiltonian completions \cite{Wu2005,Gamarnik2005}
(i.e., adding the minimum number of edges to a graph to make a Hamiltonian path
that visits each vertex once) and minimal Chordal completions \cite{Parra1997}
(i.e., adding the minimum number of edges to a graph such that every
sufficiently long cycle has a chord).  However, with the advent of effective and
scalable tools for \emph{large scale low-rank matrix} completion, in our work we
choose to focus on matrix completion techniques that make a \emph{low-rank
assumption}.

Of course, given the natural connection between adjacency matrices and graphs,
one would be tempted to look at low-rank structures in matrices such as $A$.
Much is known in such cases, including the facts that \cite{Gutman2011}:

\begin{itemize}
  \item The only rank-0 adjacency matrix is for the graph with no edges.
  \item The only rank-1 adjacency matrix is the complete graph.
  \item The only rank-2 adjacency matrix is the complete bipartite graph (and
  complete tripartite is rank-3, etc.).
\end{itemize}
\noindent It is also known, for example, that for a subgraph $S \subseteq G$
$rank(S) \le rank(G)$ \cite{Gutman2011}.

Note, \emph{Graph Laplacians} is an area in which the low-rank structure of graphs can be
precisely defined and is well understood
\cite{Chung1996}. 
We merely observe that given an
adjacency matrix $A$ one can define the diagonal matrix $D_{row}$ as the row
sums of $A$.  Given such a $D_{row}$, a Graph Laplacian for $A$ is defined
as $L=D_{row}-A$. It is well known that the Graph Laplacian is low-rank if, and
only if, the underlying graph is disconnected \cite{Chung1996}.   Unfortunately,
such techniques do not lead to the types of predictions that we desire.
\subsection{Low Rank Structure of EDMs}
Moving away from graphs for 
a moment, we observe that there is a domain in
which low-rank matrix completion has been used successfully for many years, and
this is in the completion of EDM
\cite{Gower1982,Krislock2012}. 
There is a vast literature on such matrices, and especially their
low-rank structure.  In particular, for a EDM $D$ one can show that $rank(D) \le
k+2$ where $k$ is the dimension of the embedding space of the points whose pairwise distances comprise $D$ \cite{Krislock2012}. 

Unfortunately, one is then left with the task of \emph{defining and measuring
these Euclidean distances}, which can be a non-trivial task.  As observed
previously, the relationship between the \emph{communication distance} and the
\emph{EDM} between nodes in a network can be quite
complicated, especially in the presence of routing algorithms, occlusions, and
anything but the most trivial spatial geometry of the sensors.   \emph{Just
because the sensors can physically be embedded in a 2-D or 3-D Euclidean space
does not imply that the distances implied by the communication network can be
embedded in the same space.} Accordingly, 
a goal in the current text is to
understand what elements of the theory of EDMs can be preserved while having the
results be applicable to large scale networks.

\subsection{Exactly Low-rank HDMs}

While our focus is on real-world networks, and calculation of the empirical
low-rank structure and predictability, theoretical considerations provide
important guide-posts.   For example, while the low-rankness of adjacency
matrices can be used to detect structures such as complete $l$-partite graphs
and the low-rankness of Graph Laplacians can be used to detect disconnected
graphs, we wish to treat more general scenarios.

On the other hand, the theory of the low-rank structure of HDMs provides
intriguing glimpses into what is possible when applying a low-rank assumption
to the analysis of HDMs.  For example, one has access to a vast array of
theorems of the following flavor.

\begin{theorem}

  (restated from Theorem 2.16 from \cite{Aouchiche2014a}) Let $G$ be a graph
  with HDM $H$.  If $G$ has only a single even cycle of length $2k$ and a
  total of $2k+p$ vertices, then $H$ is of rank $k$.

\end{theorem}

Interestingly, the theory of low-rank HDMs extends well beyond such special
cases.  For example, it is quite common to consider $k$-regular graphs where
each row (and column) of $A$ sum to $k$ \cite{Diestel2005}.  A similar idea,
called \emph{transmission regular}, can be defined in the HDM case by calling a
graph $G$ $k$-transmission regular when each row and column of $H$ sum
to $k$.  Such transmission regular graphs give rise to a large and interesting
class of low-rank HDMs by way of the following theorem.

\begin{theorem}

  (restated from Theorem 4.5 from \cite{Aouchiche2014a}) Let $G_1$ and $G_2$ be
  two transmission regular graphs on $n_1$ and $n_2$ vertices with transmission
  regularity $k_1$ and $k_2$, with $k_1$ and $k_2$ not necessarily equal.  If
  $G$ is the Cartesian product of $G_1$ and $G_2$, then the rank of $H$ is
  $n_1 n_2 - (n_1-1)(n_2-2)$.
\end{theorem}

\subsection{Approximately Low-rank HDMs}
Even though the above theorems give rise to interesting families of low-rank
hop-distance matrices, real-world graphs rarely, if ever, satisfy the
assumptions of these theorems, or many other similar theorems for both adjacency
and hop-distance matrices.  Accordingly, it is important to consider the
\emph{approximate low-rank properties} of hop-distance matrices.  For example,
one can consider common synthetic models that approximate the structure of
real-world graphs such as \emph{scale-free networks} 
or \emph{power-law graphs}, whose degree distributions follow, at least
asymptotically, a power-law.  For example, one often considers graphs where the
fraction of nodes $N_k$ having $k$ links goes like $N_k \propto k^{-\gamma}$
for some parameter $\gamma$ \cite{Holme2002,Barabasi1999}.

\begin{figure}[ht]
\begin{center}
\includegraphics[width=4.0in]{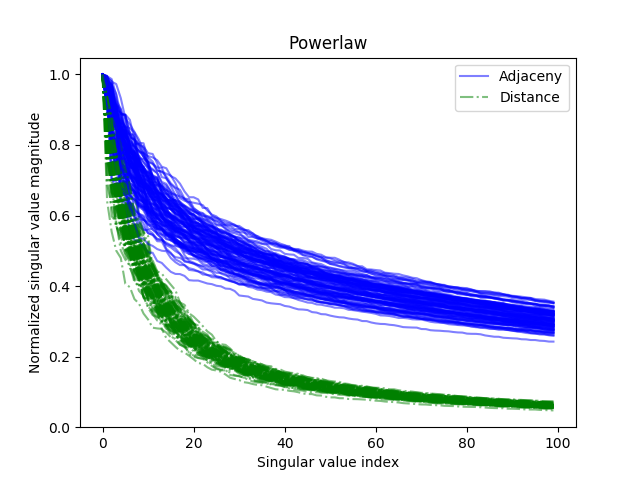}\\
\includegraphics[width=4.0in]{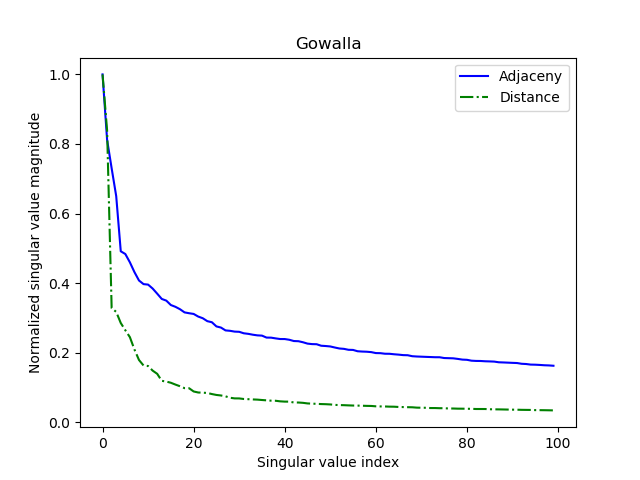}\\
\end{center}
\caption{
Comparison of the normalized singular values of
adjacency matrices versus hop-distance matrices for synthetic and
real-world power-law graphs. Top figure is for a synthetic power-law graph
generated using the Holme-Kim model \cite{Holme2002} with 500 nodes and bottom figure
is for 2000 node subgraph of the real-world Gowalla social network from
\cite{Leskovec2014}. 
}
\label{powerlaw}
\end{figure}

Accordingly, in Figure~\ref{powerlaw}, we show a comparison between the singular
values of the adjacency matrix and the hop-distance matrix for two graphs,
namely a synthetic power-law graph based upon the Holme-Kim model \cite{Holme2002} with 500 nodes and a subgraph of the real-world Gowalla social
network from \cite{Leskovec2014} with 2000 nodes.  In addition, for the
Holme-Kim synthetic data \cite{Holme2002} we examine 100 Monte-Carlo runs to see the difference
that the precise state of the random graphs makes. Note, in both cases the
matrices are double-centered (as discussed in Section~\ref{nonlinear
dimension reduction}) and the value of all singular-values are normalized relative to
the size of the largest singular value (so all curves start on the left at 1).
In both cases, the singular-values of $H$ 
decay much
more quickly than the singular-values of $A$, 
with the
normalized 100th singular value of $H$ in both cases being very close to 0, while the
same for 
$A$ are approximately 0.4
and 0.2 respectively. 

\begin{figure}[ht]
\begin{center}
\includegraphics[width=4.0in]{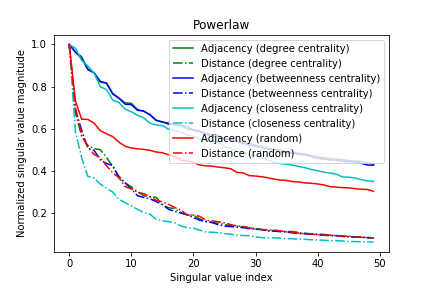}\\
\includegraphics[width=4.0in]{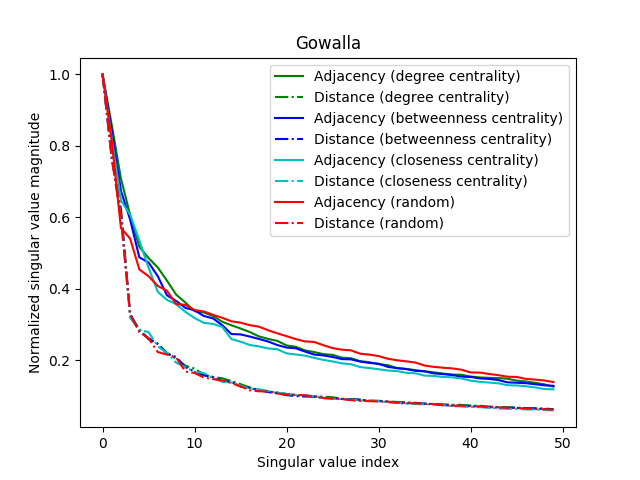}\\
\end{center}
\caption{
Comparison of the normalized singular values of
adjacency matrices versus hop-distance matrices for a variety of anchor selection strategies for 100 anchor nodes.  Top figure is for a synthetic
power-law graph generated using the Holme-Kim model \cite{Holme2002} with 500
nodes. Bottom figure is for a subgraph of the real-world Gowalla social
network from  \cite{Leskovec2014} with 2000 nodes.  
}
\label{centrality}
\end{figure}

Finally, 
we consider how the placement of anchor nodes affects
the low-rank structure of the HDM.  
In Figure~\ref{centrality} we
compare the low-rank structure of the HDM for the Holme-Kim \cite{Holme2002} and
Gowalla social network \cite{Leskovec2014} between choosing a random set of
anchors of $H$ versus choosing a set of anchors based on different 
centrality measures.  Somewhat surprisingly, the low-rank structure of $H$ is
not strongly affected by the choice of the sampling scheme.   
In particular,
Figure~\ref{centrality} suggests that the accuracy of predictions from random
anchors will be similar to that from anchors with high centrality (e.g., using nodes with a large number of neighbors as anchors) \cite{Diestel2005}.
Note that the difference between the singular values of the hop-distance and
adjacency matrices is much larger than the difference between the various
sampling schemes.  Of course, this does not mean that specially chosen anchors
cannot change the low-rank structure of $H$, but it does suggest that 
random anchors is a reasonable anchor selection for initial investigation. 

%
%
%
%
%

%% file: approach.tex
\section{Approach}
\label{approach}
Proposed method for extracting network topology  is as follows: 
\begin{itemize}
	\item Start with a set of geodesics.  Two sampling schemes are considered: anchor-based VCs (to obtain $P$ or a subset thereof)  and random shortest paths (to obtain a subset of elements of $H$).
	\item Complete the matrix $P$ or $H$  using low-rank matrix completion.
	\item Evaluate the accuracy of the computed topology or layout. In case of 2-D and 3-D sensor networks the accuracy of resulting topology preserving maps is used as the evaluation metric, while for the social networks  we evaluate the difference between the actual and completed distance matrices. 
	%
	
\end{itemize} 
Here we describe the different elements of our approach in more detail. 

\subsection{Anchor Based VCs}

We follow the notation in \cite{Jayasumana2016a} and consider networks where $M$
of the $N$ nodes are designated as ``anchors''. With an anchor-based VCS, each of the $N$ nodes in the network is characterized
by a VC vector of length $M$, i.e.,  each node is labeled by its
shortest-path hop distance to each of the $M$ anchors.

Let $P\in\mathbb{N}^{N \times M}_0$ be the matrix containing the VCs of all the
nodes, e.g., the $i$-th row corresponds to the $\mathbb{N}^{1 \times M}_0$ VC
vector of the $i$-th node, and $j$-th column corresponds to the $M$-th virtual
coordinate of all the nodes in the network with respect to $j$-th anchor.

This matrix can be written as

\begin{equation}
P =
\begin{bmatrix}
h_{1 A_1} & \hdots & h_{1 A_M} \\
\vdots & \ddots & \vdots \\
h_{N A_1} & \hdots & h_{N A_M} \\
\end{bmatrix}
\label{P definition}
\end{equation}
\noindent where $h_{i A_j}$ is the hop-distance from node $i$ to anchor
$A_j$.  $P$ is precisely a subset of the full hop distance matrix
$H$ derived by selecting just a few anchor nodes and constructing $P$
from the columns corresponding to those anchor nodes.  For large networks
it is generally desirable to have only a small
subset of nodes as anchors, i.e., $M \ll N$.

In particular, one can equivalently think of $P$ as a (non-principal)
\emph{sub-matrix} of the full hop-distance matrix $H$.  If we decompose $H$ into
blocks by writing

\[
H =
\begin{bmatrix}
A & B^T \\
B & C
\end{bmatrix},
\label{Hdecomposed}
\]

\noindent then $A \in \mathbb{N}^{M \times M}_0$ contains the hop-distances
between the $M$ anchors and themselves, $B \in \mathbb{N}^{(N-M) \times M}_0$
contains the hop-distances between the $M$ anchors and the $N-M$ non-anchor
nodes, and $C \in \mathbb{N}^{(N-M) \times (N-M)}_0$ contains the hop-distances
between the $N-M$ non-anchor nodes and themselves, which in our case are missing
entries.  In this way, $P$ can equivalently be written as

\[
P =
\begin{bmatrix}
A \\
B
\end{bmatrix}.
\label{Pdecomposed}
\]

\noindent Note, the prediction of both $P$ and $H$ from partial observations are
of interest in the current context.  Accordingly, in Section~\ref{results}, we
will provide numerical results for both problems.

\subsection{Low Rank Matrices}
\label{Low Rank Matrices}

As can be inferred from our prior work 
\cite{Dhanapala2014}, and as we further
demonstrated above, hop distance matrices $H$ for many interesting and realistic
networks are, somewhat surprisingly, \emph{approximately low-rank}.  It is this
empirical observation that inspires our work.

A widely used tool for analyzing low-rank structure in matrices is the Singular
Value Decomposition (SVD) \cite{Eckart1936} and the closely related idea of
Principal Component Analysis \cite{bishop2006pra}.  In particular, given
our VC matrix $P \in \mathbb{R}^{N \times M}$ one can write $P$ as
\[
P = U \Sigma V^T
\]
\noindent where, assuming $M<N$ as $U \in \mathbb{R}^{N \times M}$, $\Sigma \in
\mathbb{R}^{M \times M}$, and $V \in \mathbb{R}^{M \times M}$ 
In addition, the columns of $U$ and $V$ are orthonormal (i.e., $U$
and $V$ are each sub-matrices of a \emph{unitary} matrix) and $\Sigma$ is
diagonal. Finally, the diagonal entries of $\Sigma$ are called the singular
values of $P$ and the rank of $P$ is precisely the number of non-zero singular
values.

Accordingly, one can compute an approximation of $P$ by setting the
``small'' entries of $\Sigma$ to $0$, using an appropriate threshold,
to generate an approximation $\Sigma \approx \hat{\Sigma}$.  $P$ can
then be approximated similarly by setting
$P \approx \hat{P} = U \hat{\Sigma} V^T$.  Such ideas have a long
history, with an important milestone being the seminal work of Eckart
and Young in 1936 \cite{Eckart1936}.


\subsection{Topology Coordinates and Topology Preserving Maps}

The mathematical foundation of our previous work in TPM generation from a VCS
follows from the above formulation \cite{Dhanapala2014}. Consider the principle
components of P given by,
\[
P_{SVD} = U \Sigma
\]
In the TC
 domain, each node in a 3-D network is
characterized by a triple of Cartesian coordinates $(x_{T}(i),y_T(i),z_T(i))$.
Let $[X_T,Y_T,Z_T]$ be the matrix of TCs for the entire set of nodes, i.e., the
$i$-th row is the TCs of node $i$.  Then from \cite{Dhanapala2014},

\begin{equation}
[X_T,Y_T,Z_T] = [P_{SVD}^{(2)},P_{SVD}^{(3)},P_{SVD}^{(4)}] \label{TCeq}
\end{equation}

\noindent where, $P_{SVD}^{(j)}$ is the $j$-th column of $P_{SVD}$.  Note, in the
derivation of TCs as presented in \cite{Dhanapala2014} the first singular vector
$P_{SVD}^{(1)}$ is \emph{not used} in the representation. In
Section~\ref{nonlinear dimension reduction}, we will discuss the relationship
between generating TCs without $P_{SVD}^{(1)}$ and the idea of ``double
centering'' \cite{Lee2007}.

The importance of TCs is that they capture the geometric features such as the
shape and boundaries in spite of the fact that no Euclidean distance
measurements are used. However, if some physical locations are known, then the
TCs can be transferred to approximate physical coordinates as well
\cite{Buoud2016}.

\subsection{Connections to Non-linear Dimension Reduction (NDR)}
\label{nonlinear dimension reduction}

The Topology Preserving Map (TPM) generation above is closely related to several algorithms in NDR \cite{Lee2007}.  In particular, given a \emph{squared} EDM $D$, one can
compute a ``double centering'' of $D$ 
by
writing

\begin{equation}
S = -\frac{1}{2} \left(D
- \frac{1}{N} \mathds{1} \mathds{1}^T D
- \frac{1}{N} D \mathds{1} \mathds{1}^T
+ \frac{1}{N^2} \mathds{1} \mathds{1}^T D \mathds{1} \mathds{1}^T\right)
\label{double centering}
\end{equation}

\noindent where $\mathds{1} \in \mathbb{R}^{n \times 1}$ is the vector all of
whose entries are $1$ \cite{Lee2007}.  In effect, $\frac{1}{N} \mathds{1} \mathds{1}^T D$ is
the matrix which contains all of the column averages of $D$, $\frac{1}{N} D
\mathds{1} \mathds{1}^T$ is the matrix containing all of the row averages of
$D$, and $\frac{1}{N^2} \mathds{1} \mathds{1}^T D \mathds{1} \mathds{1}^T$ is
the matrix containing the average of all the entries of $D$.

Note that a TPM is a low-rank approximation of a double centered $S$ computed from $H$ rather than a squared EDM $D$. 
In this sense, a TPM is analogous  to the Multi-Dimensional Scaling (MDS) algorithm \cite{Kruskal1964,Borg2005,Lee2007}. 

Even closer to our proposed technique, others have considered \emph{geodesic}
distance matrices $D_G$ \cite{Tenenbaum2000,Lee2007} generated by drawing short
range distances from $D$, say by using a proscribed number of neighbors or only
considering distances below a certain threshold, but computing the rest of the
distances by other means.  In particular, one computes $D_G$ by selecting some
number of neighboring points for each point $x$ (e.g., all of the points laying
in some $\epsilon$-ball around $x$) and then completing $D_G$ by computing
shortest-paths in the resulting weighted graph.  A low-rank approximation of
such a geodesic based distance matrix $D$ (after double centering as in
\eqref{double centering}) is equivalent to the \emph{Isomap}
\cite{Tenenbaum2000,Lee2007} algorithm for NDR.

Intuitively, Isomap can be thought of as a relaxation of MDS to the case where
Euclidean distances are ``trusted'' for short range interactions, but not
``trusted'' for long range interactions.  
The long range
interactions are instead approximated by geodesic distances, which are thought
to be more faithful to the true geometry and topology of the network.  
Our
method generalizes this argument by assuming that not even short range
distances are to be ``trusted'' and instead our HDM is computed from unweighted
connectivity information.
Accordingly, our method uses geodesic distance
matrix $H$ analogous to $D_G$ in Isomap.  \emph{However, all of our short range
distances are presumed to be $1$, } i.e., we use the number of hops.
\subsection{Matrix Completion}
\label{matrix completion}

Prior work on TPM generation is based on the case where entire columns are taken
from $H$ and used to construct $P$.  However, in the current work, we consider
the more interesting, and practically important case, where each anchor node
only has a \emph{partial} set of measurements to the rest of the network.
Accordingly, some entries in $P$ are \emph{not observed} and the matrix $P$ is
therefore incomplete. Predicting the unobserved entries in $P$ can be phrased as
a low-rank \emph{matrix completion} problem.   In particular, we have leveraged
modern ideas in low-rank \emph{matrix completion}
\cite{Lin2013,Candes2009,candes09ex,Paffenroth2013b,Paffenroth2012}. Space does
not afford a fulsome treatment of the theory and  implementation details for
these algorithms.   Accordingly, we merely endeavor to provide the reader with
the intuition for such approaches in the context of predicting unobserved
entries in HDMs.  

The key idea of such methods can be phrased as the following optimization
problem
\begin{equation}
L = \arg \min_{L_0} \rho(L_0), \ \ 
\text{s.t.}\; \mathcal{P}_{\Omega}(M) = \mathcal{P}_{\Omega}(L_0)\label{MC} \\
\end{equation}
\noindent where $M$ is an arbitrary matrix, $\rho$ is the rank operator and
$\mathcal{P}_{\Omega}$ is an operator that extracts from $M$ the set of observed
entries designated by $\Omega$ (i.e., the constraint in (\ref{MC}) is only enforced
at the observed points). In other words, we seek to find a matrix $L_0$ such that the
rank of $L_0$ (denoted $\rho(L_0)$) is minimized while enforcing the constraint
that the matrix we construct matches our observed entries
$\mathcal{P}_{\Omega}(M)$.  Since, we enforce the constraint that
$\mathcal{P}_{\Omega}(M) = \mathcal{P}_{\Omega}(L_0)$ the returned matrix $L_0$
will be faithful to our measured hop-distances but $L_0$ is free to take on any
values it likes outside of $\Omega$ to minimize its rank.

Unfortunately, as stated, \eqref{MC} is an NP-hard optimization
problem, and can only be solved for small networks.  Recent results
\cite{Lin2013,Candes2009,candes09ex,Paffenroth2013b,Paffenroth2012}
allow, under mild assumptions, for the NP-hard optimization in
\eqref{MC} to be recast as a convex optimization problem
\begin{equation}
L = \arg \min_{L_0} \| L_0 \|_*, \ \ 
\text{s.t.}\; \mathcal{P}_{\Omega}(M) = \mathcal{P}_{\Omega}(L_0)\label{MCconvex} 
\end{equation}

\noindent where $\| L_0 \|_*$ sum of the singular values of $L_0$, often called
the \emph{nuclear-norm} of $L_0$.  The optimization problem in \eqref{MCconvex}
is \emph{convex} and can easily be solved for millions of nodes on commodity
computing hardware using splitting techniques and iterative matrix decomposition
algorithms \cite{Lin2013,Paffenroth2012,Paffenroth2013b}.  

%
%
%
%

\subsection{Completion of Partially Observed Hop-Distance Matrices}

Simply stated, our proposed method for computing VCs from partially observed
HDMs revolves around combining the NDR ideas from
Section~\ref{nonlinear dimension reduction} with the matrix completion ideas
from Section~\ref{matrix completion}.  However, one impediment remains, namely,
the double centering operation in \eqref{double centering}, {\it prima faci},
would seem to require a fully observed matrix $H$, 
negating our ability to analyze partially observed HDM matrices.

However, this difficulty in computing a ``double centering'' of a partially
observed $P$ can be overcome by way of the following equation,
similar to Equation (\ref{double centering}), 
\begin{equation}
S_{i,j} = -\frac{1}{2}
\left(P^2_{i,j} -
\mu_j (P^2) -
\mu_i (P^2) +
\mu_{i,j} (P^2)\right)
\label{double centering partial}
\end{equation}

\noindent where $\mu_j(P)$ is the mean of the observed entries in the $j$-th
column of $P$, $\mu_i(P)$ is the mean of the observed entries in the $i$-th row
of $P$, and $\mu_{i,j}(P)$ is the  mean of all of the entries in $H$. In effect,
each entry of the double-centered matrix $S_{i,j}$ only depends on the square of
the single entry $P_{i,j}$, along with \emph{mean values} of the \emph{rows} and
\emph{columns} of $P$.  Accordingly, estimates of these mean values can be
computed even for a partially observed matrix such as $\mathcal{P}_{\Omega}(P)$,
by performing the required mean over just the observed entries of the
appropriate column, row, or the entire matrix. 
Of course, if a particular node has no measurements such a node cannot 
be predicted. 
Also, classically, \eqref{double centering partial} is defined
for EDMs, and one might wonder if it is applicable to HDMs?  In fact, that is
a salient point of our work, to find a space in which the
Euclidean-distances and the hop-distances coincide.

In some sense, such restrictions on acceptable sampling schemes for $H$ should
not be surprising.  For example, if a particular row of $P$ (or $H$) contains no
measurements then the distance from this node to any anchor is not known. In
effect, nothing is known about this node and therefore no predictions can be
made.

Identifying classes of acceptable and unacceptable sampling schemes is a topic
for future research. However, such ideas are a close cousin of the incoherence
requirements that arise in matrix completion problems
\cite{Lin2013,Candes2009,candes09ex,Paffenroth2013b}.  Accordingly, drawing
inspiration from that literature and for simplicity, we choose \emph{random
nodes as anchors} and \emph{random nodes} whose distance we measure from each
anchor.

%

Our algorithm for recovery of complete $P$ from partial entries and generating
TCs from a partially observed HDM is described in
Procedure~\ref{completeHDMalgorithm} below.
Topology coordinates in \cite{Dhanapala2014}  are given by 2nd and 3rd singular vectors in case of 2-D networks and 2nd, 3rd and 4th in case if 3-D networks as given in Equation~\ref{TCeq}. Thus as a comparison we use  Procedure~\ref{completeHDMalgorithmP}, which carries out matrix completion directly on $P$, and follows the approach in \cite{Dhanapala2014} to generate TPMs.  
Procedure~\ref{completeHDMalgorithm}, based on double centering followed by the completion of the Grammian matrix $S$, however,  follows the approaches such as MDS for NDR in Euclidean spaces more closely. 
\begin{algorithm}
	\begin{algorithmic}[1]
		\Require Partially complete $P$ of a graph $G=\{V,E\}$
		\Require A target dimension $k$
		\Procedure{Complete $S$ from Partial $P$}{}
		\State Compute $\mathcal{P}_{\Omega}(S)$ from $\mathcal{P}_{\Omega}(P)$
		using \eqref{double centering partial}
		\State Compute approximate Grammian matrix $S$ from $\mathcal{P}_{\Omega}(S)$ using \eqref{MCconvex}
		\State Compute the SVD $S = U \Sigma V^T$
		\State \textbf{return} The first $k$ columns of $U \Sigma$ are the TCs
		\EndProcedure
	\end{algorithmic}
	\caption{Computing TPMs from a partially observed HDM matrix $P$ via completion of Grammian matrix.}
	\label{completeHDMalgorithm}
\end{algorithm}
\begin{algorithm}
	\begin{algorithmic}[1]
		\Require Partially complete $P$ of a graph $G=\{V,E\}$
		\Require A target dimension $k$
		\Procedure{Complete $P$ from Partial $P$}{}
		\State Compute approximate distance vector matrix $P$ from $\mathcal{P}_{\Omega}(P)$ using \eqref{MCconvex}
		\State Compute the SVD $P = U \Sigma V^T$
		\State \textbf{return} Columns $2$ through $k+1$ (i.e., the first column is excluded) are the TCs
		\EndProcedure
	\end{algorithmic}
\caption{Computing TPMs from a partially observed HDM matrix $P$ via matrix completion.}
	\label{completeHDMalgorithmP}
\end{algorithm}  

%% file: results.tex
\section{Results}
\label{results}

Here, we demonstrate the effectiveness of the proposed HDM based approach described in Procedure~\ref{completeHDMalgorithm} in
constructing accurate topology maps from a small set of
hop-distances among node pairs.  In particular, we demonstrate how $P$ or $H$ can be
recovered from a set of partial observations, and how topological coordinates
that arise from the eigen-decomposition of $P$ provide accurate recovery
of relationships among network nodes.  The evaluation is carried out
for a set of 2-D and 3-D sensor networks, and in Appendix \label{FirstAppendix} also for a social network
representing classes of networks where physical coordinates play no role.
All the results have been averaged over 100 experimental iterations.
As these results demonstrate, our methods provide surprisingly accurate
predictions, even when only a tiny fraction of the network has been measured.

\subsection{Recovery of Networks Embedded in 2-D/3-D Spaces}
\label{recoverP}
\begin{figure}[ht]
\begin{center}
\includegraphics[width=4in]{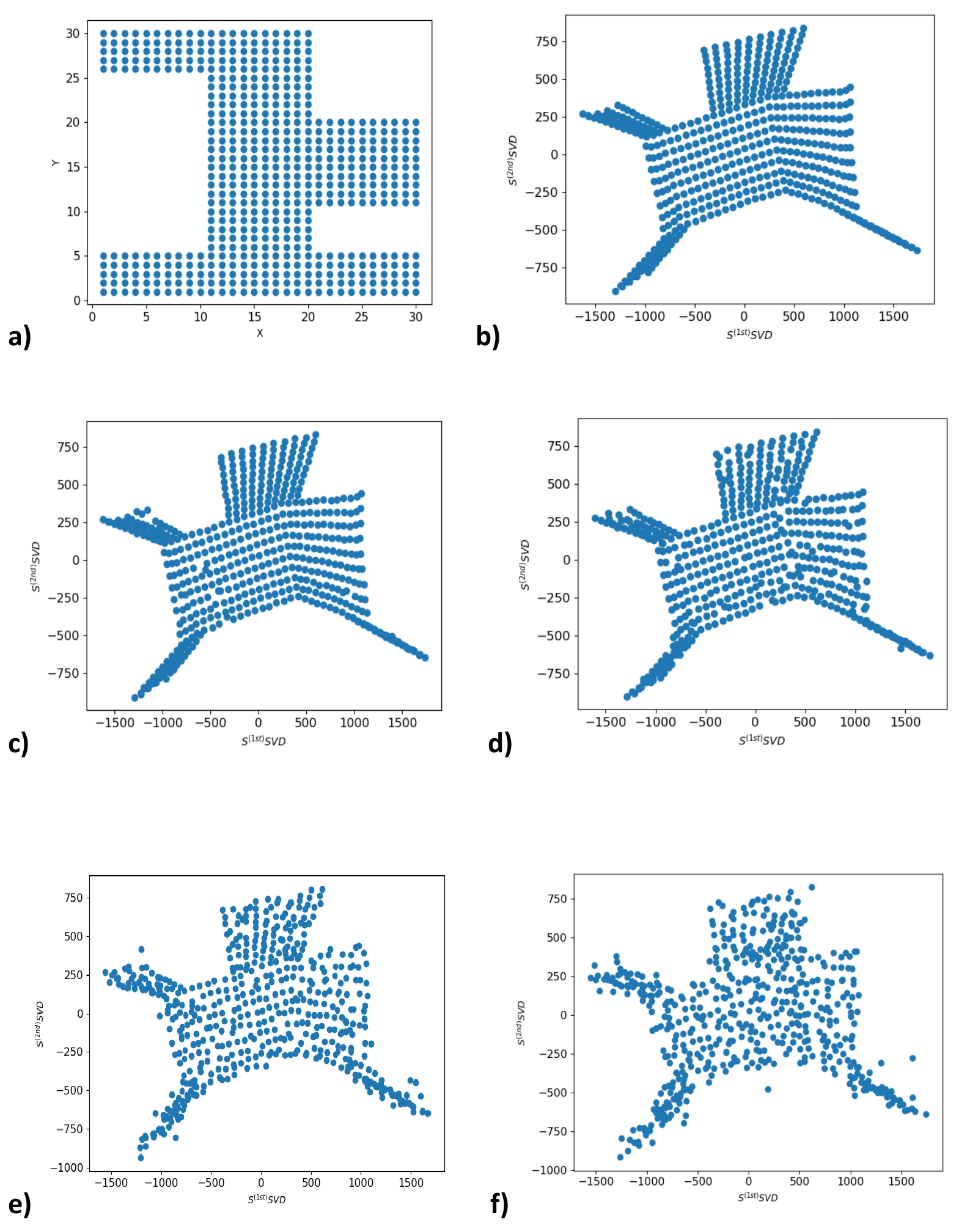}
\end{center}
\caption{
Concave network: (a) Original layout, and (b) TPM recovered from full set of VCs
with 20 random anchors;  Recovered TPM with (c) 10\%, (d) 20\%, (e) 40\%, and
(f) 60\% of sampled coordinates randomly discarded.  
}
\label{concave}
\end{figure}

\begin{figure}[ht]
\begin{center}
\includegraphics[width=4in]{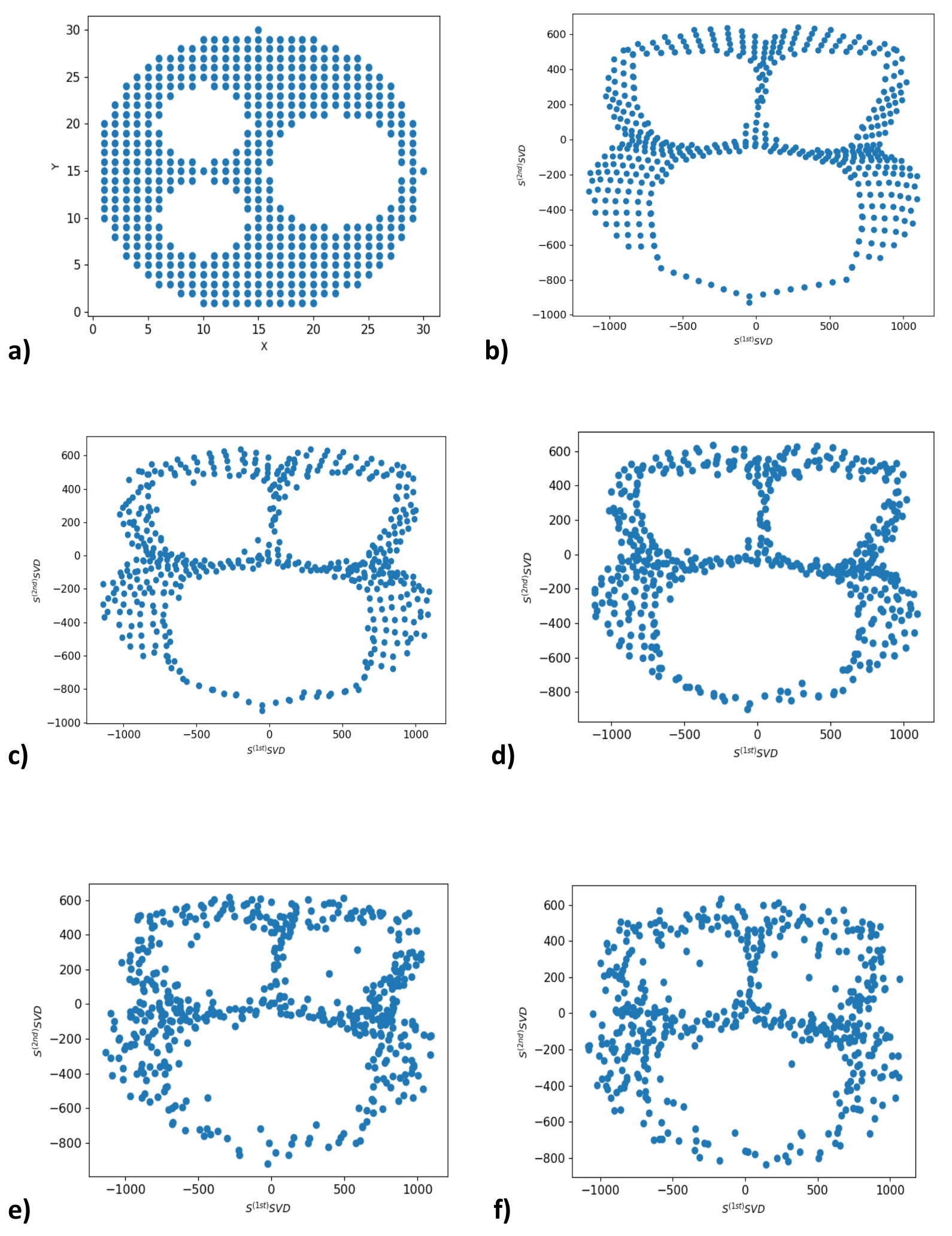}
\end{center}
\caption{
Circular network: (a) Original layout, and (b) TPM recovered from full set of
VCs with 20 random anchors;  Recovered TPM with (c) 10\%, (d) 20\%, (e) 40\%, and
(f) 60\% of sampled coordinates randomly discarded.
}
\label{circular}
\end{figure}

\begin{figure}[ht]
\begin{center}
\includegraphics[width=4in]{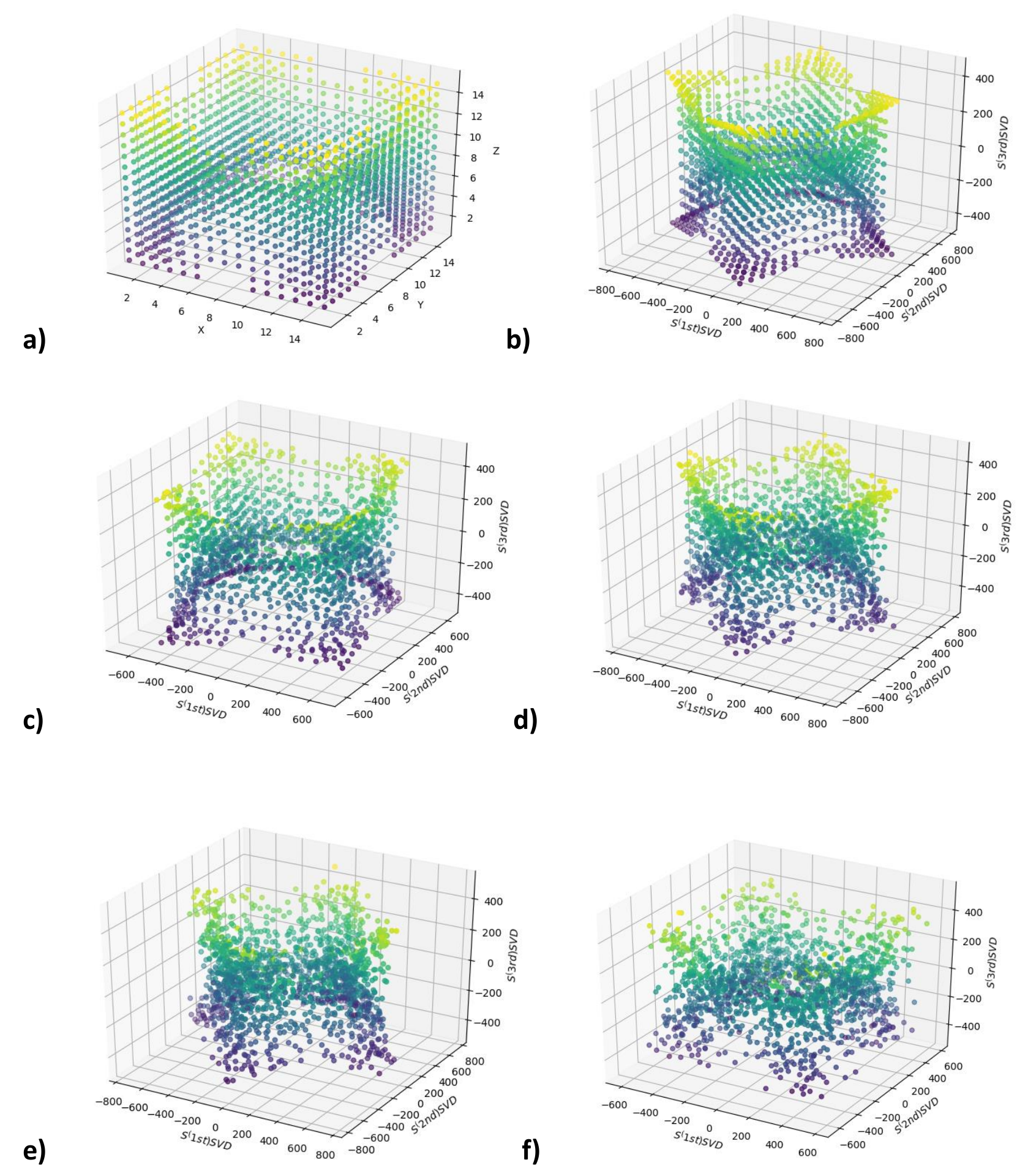}
\end{center}
\caption{
Cube with hourglass shaped void: (a) Original layout, and (b) TPM
recovered from full set of VCs with 20 random anchors;  Recovered TPM with (c)
10\%, (d) 20\%, (e) 40\%, and (f) 60\% of sampled coordinates randomly discarded.
}
\label{cube}
\end{figure}

\begin{figure}[ht]
\begin{center}
\includegraphics[width=4in]{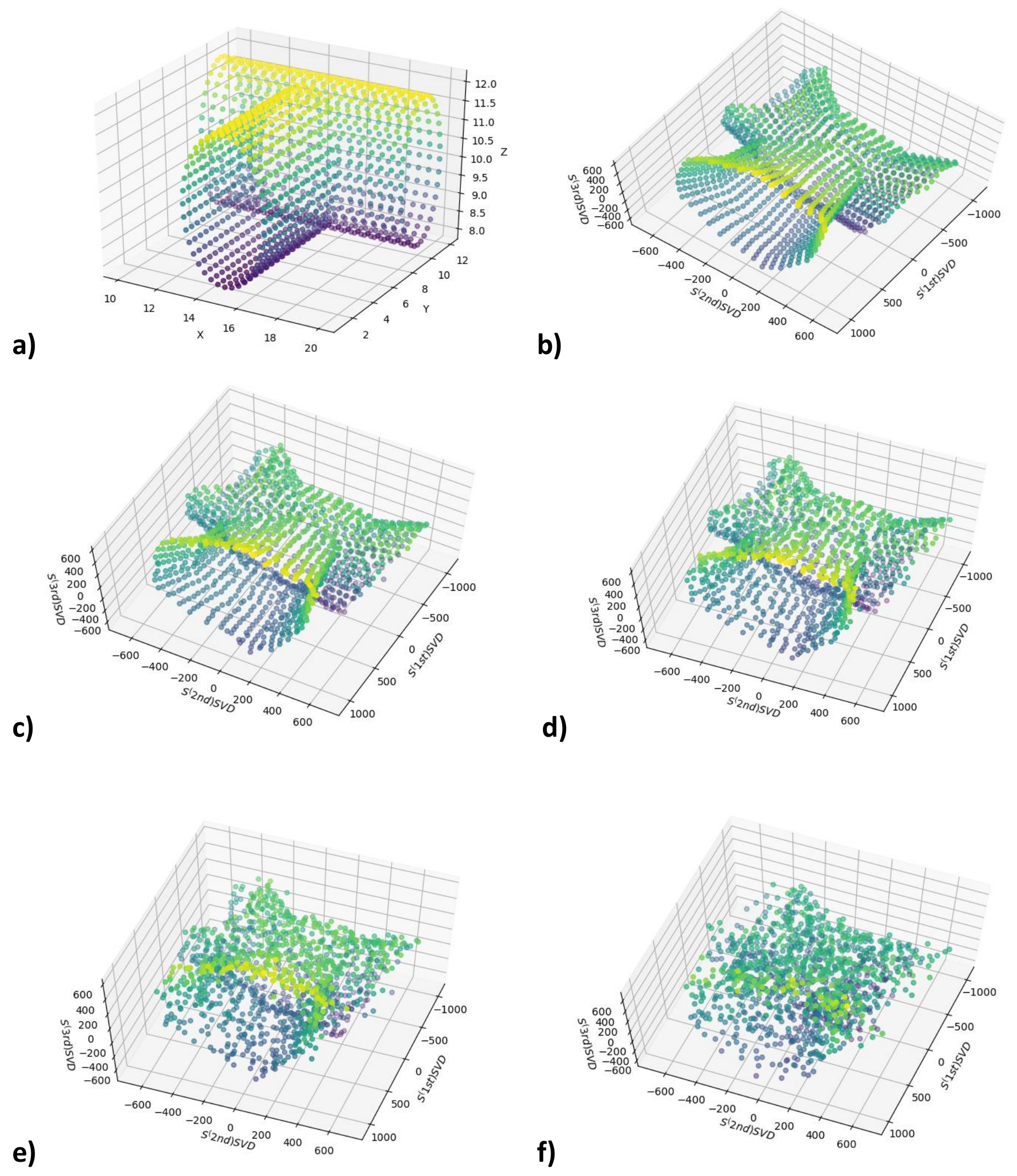}
\end{center}
\caption{
Hollow T shaped Cylinder: (a) Original layout, and (b) TPM recovered from full
set of VCs with 20 random anchors;  Recovered TPM with (c) 10\%, (d) 20\%, (e)
40\%, and (f) 60\% of sampled coordinates randomly discarded.
}
\label{t shaped}
\end{figure}

Four networks representative of 2-D and 3-D sensor network deployments covering a range of shapes and sizes are used for the evaluation. They contain complex features such as convex and concave boundaries and voids.
\begin{itemize}

\item A concave 2-D network with 550 nodes, the physical layout of which
is shown in Figure~\ref{concave}(a) \cite{Dhanapala2014}.

\item A 2-D circular network with multiple circular voids of
496 nodes as shown in Figure~\ref{circular}(a) \cite{Dhanapala2014}.

\item A 3-D network, shown in Figure~\ref{cube}(a), consisting of
1640 nodes, which occupies a cube shaped volume with a hollow region in the
shape of an hourglass devoid of nodes \cite{Jayasumana2016a}.

\item A 3-D surface network, shown in Figure~\ref{t shaped}(a),
consisting of 1245 nodes, which is comprised of two hollow cylinders joined in a
``T'' configuration \cite{Dhanapala2014}.

\end{itemize}
%

\begin{figure}[ht]
\begin{center}
\includegraphics[width=4.0in]{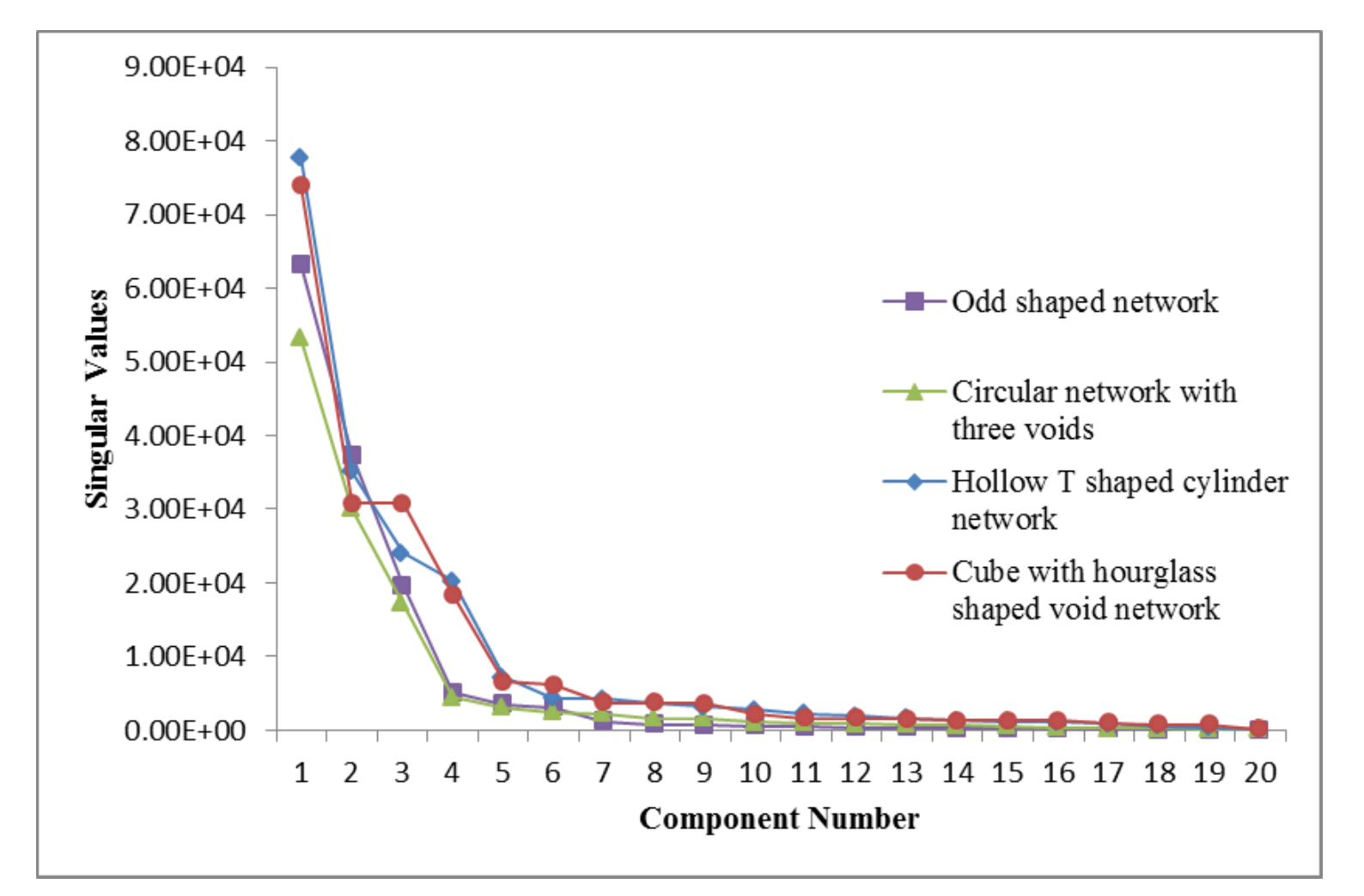}
\end{center}
\caption{
Singular values of the VC matrix for  Circular Network,  Odd-Shaped
Network, Hollow T cylinder network, and the 3-D network with void indicating the
low-rankness of VCS data.
}
\label{singular values}
\end{figure}

As an initial exploration of the applicability of the techniques we propose, we
first examine the rank of the VC matrices ($P$) for each of our networks. Twenty
random anchors were selected in each case, i.e., $M=20$, which corresponds to
approximately $3.6\%$, $4\%$, $1.2\%$, and $1.6\%$ of the nodes, respectively,
for each of our four networks. The singular values of the full VC matrices of
the four networks are shown in Figure~\ref{singular values}.  If we were
considering EDMs, then the rank of the first two networks would be $4$ (since
they are embedded in $\mathbb{R}^2$) and the rank of the second two networks
would be $5$ (since they are embedded in $\mathbb{R}^3$)
\cite{Gower1982,Krislock2012}.  As seen in Figure~\ref{singular values},
the rank of the HDMs is certainly higher than their embedding dimensions would
indicate, if they were EDMs.  However, somewhat surprisingly, even though the
four networks are quite different, all of their ranks are substantially smaller
than $20$, for the chosen random anchors.  Our interest is in the recovery of
topological information and geometric relationships such as the general shapes
of boundaries, voids in the networks, and node neighborhood preservation. Thus,
the question is whether such information is preserved and can be extracted from
small numbers of anchors and partial observations of $P$.

Two-dimensional TPMs extracted using the full set of VCs following
\cite{Dhanapala2014}  are shown in Figure~\ref{concave}(b),
Figure~\ref{circular}(b), Figure~\ref{cube}(b), and Figure~\ref{t shaped}(b),
respectively, for the four networks. It is important to note that even the full
set of VCs corresponding to 20 anchors, which corresponds to 20 random columns
of $H$, contains only  approximately $3.6\%$, $4\%$, $1.2\%$, and $1.6\%$
elements of the corresponding HDM.

Next, we randomly discard  10\%, 20\%, 40\% and 60\% respectively of this
already small sample of the elements of $H$.  The TPMs recovered using low-rank matrix
completion followed by TPM extraction are shown in Figure~\ref{concave}(c-e),
Figure~\ref{circular}(c-e), Figure~\ref{cube}(c-e), and Figure~\ref{t
shaped}(c-e) for these various sub-samplings. The results indicate that accurate TPMs
of networks are obtainable with only a fraction of virtual coordinates. 
It is important to
recognize that the goal of this work is not to necessarily recover the
maps in subfigures (a) of Figures ~\ref{concave}-\ref{t shaped}.  Rather,  we wish to recover subfigures (b) (the fully observed
topology map) from a sparse set of hop-distance observations.  


\begin{figure}[ht]
\begin{center}
\includegraphics[width=4in]{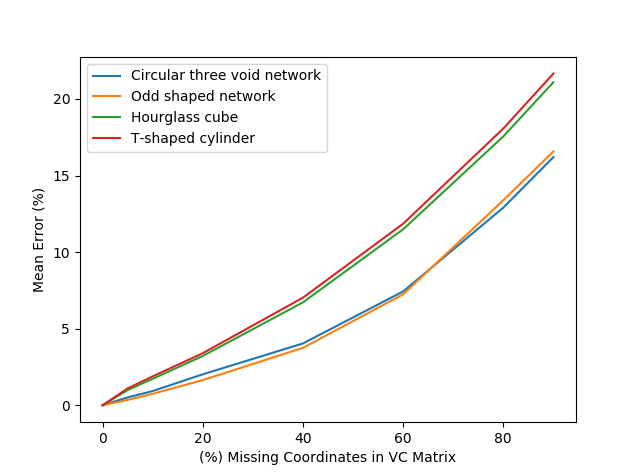}
\end{center}
\caption{
Mean error [as defined in \eqref{equation mean error}]
versus the percentage of missing virtual coordinates for sensor networks.
} \label{mean error} \end{figure}

To precisely quantify the error introduced to the TPM due to missing VCs, we
define the mean error $E$ as follows:

\begin{equation}
E = { \left[ \sum\limits_{i=1}^N \sum\limits_{j=1}^M  {| d_{ij}(f)- d_{ij}(0) |} \right]}\left/{
	\left[\sum\limits_{i=1}^N \sum\limits_{j=1}^M  {d_{ij}(0)} \right]}\right.,
\label{equation mean error}
\end{equation}

\noindent where, $d_{ij}(f)$ refers to the Euclidean distance between nodes $i$
and $j$ on the TPM when $f$ fraction of random anchor coordinates are missing.
The percentage mean  error with percentage of missing VCs for
the four networks are shown in Figure~\ref{mean error}. It is important to note
that even when mean error is high, much of the local neighborhood and shape
information is preserved. 
\subsection{Accuracy of 2-D Topology Preservation}
The accuracy of neighborhood preservation of reconstructed topology maps of  2-D networks is evaluated below using the topology preservation error $E_{TP}$, which captures the degree to which  the neighborhood relationships are altered. We provide only a brief explanation of $E_{TP}$ and refer the  reader to \cite{Dhanapala2014} for its precise definition. Consider the network in Figure~\ref{concave}(a). The network is scanned along the set of lines in the horizontal direction ($\vec{H}$) and the vertical direction ($\vec{V}$).  Let  $\alpha$ and $\beta$  denote the sets of lines in $\vec{H}$ and $\vec{V}$ directions respectively.  Consider one such line which contains the  ordered set of some  $m$ nodes $\{n_1,... n_m\}$, which we call the original placement.   Now consider the projection of this set of nodes in a TPM (e.g., Figure~\ref{concave}(b)) on the corresponding axis. An 
error indicator function $I_{i,j}$ is defined by comparing the order of projected nodes with the original placement as: 
\begin{equation}
I_{i,j} =
\begin{cases}
1, & \text{nodes $i$ and $j$ are out of order,} \\
0, & \text{nodes $i$ and $j$ are in the same order or $i=j$.} \\
\end{cases}
\label{indicator}
\end{equation}
For the line under consideration, the neighborhood preservation error is quantified by 
$
\small
{  \sum\limits_{\forall i,j}  (I_{i,j}) }/{{^{m}{P}_2}},
$
where ${^{m}{P}_2}  $ is permutation of $m$ objects taken 2 at a time. 
However, in case of TPM, we are interested in the overall neighborhood preservation error $E_{TP}$ over the set of lines in $\vec{H}$ and $\vec{V}$ directions, which is given by \cite{Dhanapala2014}: 
\begin{equation}\small
E_{TP} =
{ \left[ \sum\limits_{\alpha}  \sum\limits_{\forall i,j}  (I_{i,j}) + \sum\limits_{\beta}  \sum\limits_{\forall i,j}  (I_{i,j}) \right]}\left/\left[{\sum\limits_{\alpha}{^{m}{P_2}}} + \sum\limits_{\beta}{^{m}{P_2}}\right]\right.
\label{TPM-error-o}
\end{equation}

Next we evaluate the effectiveness  of matrix completion with HDMs for TPM generation using the approaches in Procedure~\ref{completeHDMalgorithm} and Procedure~\ref{completeHDMalgorithmP}.  
$E_{TP}$ (as a percentage) for the 2-D networks for these two cases provided in Table~\ref{table-nw1} 
show that with merely 20 random anchors, which corresponds to only  4\% of the nodes, networks can be recovered with an error less than 5\% even  after deleting 80\% of the entries.

\begin{table}[htp]
	\small
	\centering
	\begin{tabular}{|p{15mm} p{6.4mm} p{6.4mm} p{6.4mm} p{6.4mm} p{6.4mm}|}
	\hline 
	Deletions(\%) & 10\% & 20\% & 40\% & 60\% & 80\%\\
	\hline
	$ Circular_S $  & 3.31 & 3.55 & 3.91 & 4.25 & 4.29 \\
    $ Circular_P $  & 2.47 & 2.92 & 3.73 & 4.04 & 4.30 \\
	$ Concave_S $  &3.09  & 3.47 & 3.48 & 4.43 & 4.50 \\
	$ Concave_P $  & 2.94  & 3.23 & 3.65 & 3.82 & 3.94 \\
	\hline
\end{tabular}
	\caption{Topology preservation errors ($E_{TP}\%$) 
		for circular and concave networks with  10\% to 80\% of sampled coordinates randomly discarded. Subscript $S$ denotes results with completion of Grammian matrix and taking 1st and 2nd singular vectors (Procedure~\ref{completeHDMalgorithm}), while subscript $P$ denotes completion of the $P$ matrix and taking 2nd and 3rd singular vectors  (Procedure~\ref{completeHDMalgorithmP}). }
	\label{table-nw1}
\end{table}

%% file: conclusions.tex
\section{Conclusion}
\label{conclusion}
This paper addresses the problem of recovering network features from a small
set of hop-based graph geodesics.
For networks deployed on 2-D surfaces and 3-D
spaces the geometric and physical layout features are of importance.  The
approach starts with anchor-based VCs but, unlike prior techniques that required
the entire VC set, the proposed approach requires only a fraction of the
measured VCs to recover accurate topology preserving maps. Our technique is
based on the theory of low-rank matrix completion that reconstructs missing VCs,
the result of which is used to recover layout maps using topology preserving map
generation techniques.  The results presented here not only allow the reduction of cost (communication,
power, etc.) of VC generation but, more importantly, open the possibility of
using topology coordinate based techniques for large networks and even those
involving soft-state systems, where some coordinate values may be allowed to
expire, thus allowing for more resilient network operations.

With random anchor selection, the VC matrix ($P$) is equivalent
to a small set ($<10$\%) of random columns of an HDM.  Thus, the partially complete
$P$ matrix where  a large set of random VC entries are missing is equivalent to
an incomplete HDM with only a very small number of entries.
Beyond results for physically embedded networks, we also demonstrated the ability to make accurate predictions in a large social network also from a small set of random geodesic measurements. 
We also demonstrated that the HDMs of many  real-world networks are low-rank. Therefore, the approach presented here provides a foundation for designing novel graph sampling techniques that 
allows the capture of complex real-world networks with a small number of measurements. 

%% file: appendix.tex
\appendix
\label{FirstAppendix}
Note: The material in this  Appendix is included for reviewer information only.
It will be included in the supplementary material in the final version. 
\subsection{Table of Acronyms}
	\begin{tabular}{p{9mm} p{50mm}}
		\hline 
		Acronym & Expansion\\
		\hline
		EDM & Euclidean Distance Matrix\\
		HDM & Hop Distance Matrix\\
		MDS & Multi-Dimensional Scaling \\
		NDR & Nonlinear Dimension Reduction \\
		PCA & Principal Component Analysis \\
		SVD & Singular Value Decomposition\\
		TC & Topology Coordinates\\
		TPM &Topology Preserving Map\\
		VC & Virtual Coordinates\\
		VCS & Virtual Coordinate System\\
		\hline
	\end{tabular}
\subsection{Recovery of a Social Network}

Gowalla is a  mobile web based social media application for which the  connectivity dataset is available from the Stanford Large Network Dataset Collection \cite{Leskovec2014}. A
subnetwork of 10,000 nodes with 140,866 edges among them is used for the
evaluation.  As opposed to the results in Section~\ref{recoverP}, for this
experiment, we assume that a random set of elements of $H$ is known. 
This is equivalent to  making every node of $H$ an anchor node. \emph{However, each anchor node only measures its distance to a small set of other nodes}.

As a Euclidean distance metric, such as in \eqref{equation mean error}, is not
applicable in the case of a social network, we use mean error $E_m$
corresponding to the percentage error in the prediction of $H$, and the absolute
hop-error in geodesic lengths $E_a$ is defined as follows:

\begin{equation}\small
E_m =
{ \left[ \sum\limits_{i,j=1}^{N,N}  |\hat{h}_{ij}(f)- h_{ij}| \right]}\left/{ \left[\sum\limits_{i,j=1}^{N,N}  h_{ij} \right]}\right.,\\
\label{mean-error}
\end{equation}

\begin{equation}
E_a =
\left.{ \left[ \sum\limits_{i,j=1}^{N,N}  |\hat{h}_{ij}(f)- h_{ij}| \right]}\right/{N^2},\
\label{absolute-error}
\end{equation}

where, $\hat{h}_{ij}(f)$ refers to the element $ij$ of estimated HDM ($\hat{H}$)
when $f$ fraction of random elements are missing. Note that $\hat{h}_{ij}(0)=
h_{ij}.$

%

\begin{figure}[ht]
	\begin{center}
		\includegraphics[width=4in]{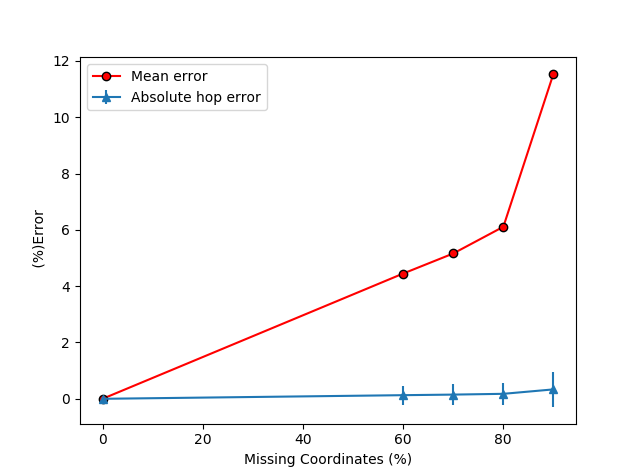}
	\end{center}
	\caption{
		Mean error and absolute hop distance error as defined in
		equations \eqref{mean-error} (in red) and \eqref{absolute-error} (in blue)
		versus the percentage of missing elements in $H$ for Gowalla social network.
	}
	\label{gowalla_error}
\end{figure}

The mean error and absolute hop distance error for different percentages of
missing elements are shown in Figure~\ref{gowalla_error}. Even with 20\% of the
elements of $H$, i.e., 80\% of coordinates missing, the network can be recovered
with an error of approximately 6\%, while the absolute hop error is less than 1.


